\documentclass[preprint,review,3p,authoryear,sort&compress]{elsarticle}
\usepackage{titlesec}
\usepackage[colorlinks=true]{hyperref}
\usepackage{parskip}
\usepackage[utf8]{inputenc}
\usepackage{verbatim}
\usepackage[skip=5pt]{caption}
\usepackage{tikz}
\usepackage{siunitx}

\usepackage{lineno}

\usepackage{algorithm}
\usepackage{algpseudocode}

\usepackage{adjustbox}
\usepackage{array}

\newcommand{\Continue}[1]{\textbf{continue on line #1}}

\usepackage{fancyhdr}
\fancypagestyle{plain}{%
  \fancyhf{}
  \fancyhead[L]{\sc Barnes}
  \fancyhead[R]{\sc Parallel Flow Accumulation}
  \cfoot{\thepage}
  
  }
\biboptions{square}

\newcommand{\todo}[1]{}

\hypersetup{
  pdfauthor   = {Richard Barnes (ORCID: 0000-0002-0204-6040)},
  pdftitle    = {Parallel Non-divergent Flow Accumulation For Trillion Cell Digital Elevation Models On Desktops Or Clusters},
  pdfkeywords = {parallel computing, hydrology, geographic information system (GIS), flow accumulation, upslope area, contributing area},
  pdfproducer = {LaTeX},
  pdfcreator  = {pdfLaTeX}
}

\makeatletter
\newenvironment{breakablealgorithm}
  {% \begin{breakablealgorithm}
   \vspace{0.5em}
   \begin{center}
     \refstepcounter{algorithm}% New algorithm
     \hrule height.8pt depth0pt \kern2pt% \@fs@pre for \@fs@ruled
     \renewcommand{\caption}[2][\relax]{% Make a new \caption
       {\raggedright\vspace{-0.6em}\textbf{\ALG@name~\thealgorithm} ##2\par}%
       \ifx\relax##1\relax % #1 is \relax
         \addcontentsline{loa}{algorithm}{\protect\numberline{\thealgorithm}##2}%
       \else % #1 is not \relax
         \addcontentsline{loa}{algorithm}{\protect\numberline{\thealgorithm}##1}%
       \fi
       \kern2pt\hrule\kern2pt
     }
  }{% \end{breakablealgorithm}
     \kern2pt\hrule\relax% \@fs@post for \@fs@ruled
   \end{center}
  }
\makeatother

\begin{document}
\thispagestyle{empty}

%\linenumbers

\begin{frontmatter}
  \title{\sc Parallel Non-divergent Flow Accumulation \\ For Trillion Cell Digital Elevation Models \\ On Desktops Or Clusters}

  \author[rb]{Richard Barnes\corref{cor_rb}}
  \ead{richard.barnes@berkeley.edu}
  \address[rb]{Energy \& Resources Group, Berkeley, USA}
  \cortext[cor_rb]{Corresponding author. ORCID: 0000-0002-0204-6040}

  \begin{abstract} %~??? words
  \noindent Continent-scale datasets challenge hydrological algorithms for processing digital elevation models. Flow accumulation is an important input for many such algorithms; here, I parallelize its calculation. The new algorithm works on one or many cores, or multiple machines, and can take advantage of large memories or cope with small ones. Unlike previous parallel algorithms, the new algorithm guarantees a fixed number of memory access and communication events per raster cell. In testing, the new algorithm ran faster and used fewer resources than previous algorithms, exhibiting ${\sim}30\%$ strong and weak scaling efficiencies up to 48 cores and linear scaling across datasets ranging over three orders of magnitude. The largest dataset tested had two trillion ($2\cdot10^{12}$) cells. With 48 cores, processing required 24 minutes wall-time (14.5 compute-hours). This test is three orders of magnitude larger than any previously performed in the literature. Complete, well-commented source code and correctness tests are available on Github.
  \end{abstract}

%Immediately after the abstract, provide a maximum of 5 keywords, avoiding general and plural terms and multiple concepts (avoid, for example, 'and', 'of'). Be sparing with abbreviations: only abbreviations firmly established in the field may be eligible. Please note that Keywords should NOT include words that already appear in the title of the manuscript. These keywords will be used for indexing purposes.
  \begin{keyword}
  parallel computing \sep hydrology \sep geographic information system (GIS) \sep upslope area \sep contributing area
  \end{keyword}
\end{frontmatter}

\section{Software}
Complete, well-commented source code, an associated makefile, and correctness tests are available at \url{https://github.com/r-barnes/Barnes2016-ParallelFlowAccum}. The code is written in C++ using MPI and constitutes 2,131 lines of code of which 58\% are or contain comments.

This algorithm is part of the RichDEM (\url{https://github.com/r-barnes/richdem}) terrain analysis suite, a collection of state of the art algorithms for processing large digital elevation models quickly.

\section{Introduction}
Digital elevation models (DEMs) are representations of terrain elevations above
or below a chosen zero elevation. Raster DEMs, in which the data are stored as
a rectangular array of floating-point or integer values, are widely used in
geospatial analysis for estimating a region's hydrologic and geomorphic
properties, including soil moisture, terrain stability, erosive potential,
rainfall retention, and stream power. Many such analyses require that every cell in a DEM have an associated flow accumulation (otherwise known as upslope area, contributing area, and upslope contributing area). Informally, if there were a rain storm, flow accumulation is directly proportional to the total amount of water which would pass through a cell as it flowed downhill from higher elevations. %TODO: More terms for flow accumulation?

DEMs have increased in resolution from 30--90\,m in the recent past to the
sub-meter resolutions becoming available today. Increasing resolution has led
to increased data sizes: current DEMs are on the order of gigabytes and
increasing, with billions of cells. Even in situations where only comparatively
low-resolution data are available, a DEM may cover large areas: 30\,m Shuttle
Radar Topography Mission (SRTM) elevation data has been released for 80\% of
Earth's landmass.~\citep{Farr2007} While computer processing and memory
performance have increased appreciably, development of algorithms suited to
efficiently manipulating large, continent-scale DEMs is on-going.

If a DEM can fit into the RAM of a single computer, several algorithms exist
which can efficiently calculate flow accumulation.~\citep{Barnes2014pf,Mark1988} If a DEM cannot fit into the RAM of a single computer, other approaches are
needed. This paper presents such an approach.

Formally, the flow accumulation $A$ of a point $p$ is defined as
\begin{equation}
\label{equ:flow_acc}
A(p)=w(p)+\sum_{n\in \mathcal{N}(p)} \alpha(n,p) A(n)
\end{equation}
where $w(p)$ is the amount of flow which originates at the cell $p$. Frequently this is taken to be 1, but the value can also vary across a DEM if, for example, rainfall or soil absorption differs spatially. The summation is across all of the cell's neighbours $\mathcal{N}(p)$. $\alpha(n,p)$ represents the fraction of the neighbouring cell's flow accumulation $A(n)$ which is apportioned to $p$. Flow may be absorbed during its downhill movement, but may only be increased by cells, so $\alpha$ is constrained such that $\sum_p \alpha(n,p)\le1\ \forall n$.

To calculate flow accumulation, a DEM is used to construct a directed acyclic graph of flow directions. The flow directions determine what fraction of the flow originating in and passing through a cell is apportioned to each of its neighbours. Though there are many ways of determining this, all flow metrics can be characterized as being either divergent or non-divergent. Non-divergent metrics, such as D8~\citep{Ocallaghan1984} and $\rho$8~\citep{Fairfield1991}, apportion a cell's flow to a single one of its neighbours. As a corollary, with such metrics two streams which join will never split apart and every cell's flow exits the DEM through a single downstream cell. Divergent methods such as D$\infty$~\citep{Tarboton1997} and MFD~\citep{Freeman1991} apportion a cell's flow to at most two and possibly many neighbours, respectively. As a corollary, with such metrics streams may bifurcate and a cell's flow may exit the DEM through many downstream cells. %There are several empirical studies which compare the results of these flow metrics.\citep{Erskine2006}  %TODO: More?
The one-to-many property of divergent flows makes developing divide-and-conquer approaches difficult, so only non-divergent metrics are considered here. Relatedly, most forms of absorption represent a simple extension of the algorithm presented here. Therefore, I consider only the case where $\alpha(n,p)=\{0,1\}$; that is, I consider only non-divergent flow metrics in which flow is directed to a single downstream neighbour.

Often, flow directions must be calculated only after internally-draining regions of a DEM called depressions (see \citet{Lindsay2015} for a typology) have been eliminated. This can be done in one of two ways. (1)~The depressions can be filled to the level of their lowest outlets. \citet{Barnes2016} discusses an efficient method for doing so on \textit{rather} large DEMs using methods based on the Priority-Flood~\citep{Barnes2014pf}. (2)~Depressions that are small or shallow enough can be breached, as in~\citet{Lindsay2015}. See \citet{Barnes2016} for a review of depression-filling in large DEMs.

In addition to depression-filling, flats (areas of a DEM with no local relief) must be assigned flow directions. This can be done by either (a)~routing flow towards only lower terrain~\citep{Domingue1988,Barnes2014pf} or (b)~routing flow both away from higher terrain and towards lower terrain~\citep{Barnes2014dd,Garbrecht1997}. Here the former option is chosen for computational efficiency. The choice of algorithms for depression filling and flat resolution do not affect any of the details of how flow accumulation is calculated.

%\section{Background}
\label{sec:alt_algs}

Existing algorithms~\citep{Gomes2012,Do2011,Yildirim2015,Arge2003,Tesfa2011,Wallis2009parallel,Danner2007,Metz2011,Metz2010,Lindsay2015,Yao2015} have taken one of two approaches to DEMs that cannot fit entirely into RAM. They either (a)~keep only a subset of the DEM in RAM at any time by using virtual tiles stored to a computer's hard disk or (b)~keep the entire DEM in RAM by distributing it over multiple compute nodes which communicate with each other. \citet{Barnes2016} reviews the designs of these algorithms and argues that both of these approaches scale poorly due to the high costs of disk access and/or communication; in contrast, the new algorithm pays much lower costs.

The algorithm presented here is superior to previous approaches because it can (a)~guarantee locality, ensuring that each DEM cell is accessed a fixed number of times, regardless of the size or content of the DEM; (b)~guarantee that all compute nodes remain fully utilised; (c)~operate using fewer nodes than would be required to hold the entire DEM; and (d)~it requires only a fixed number of low-cost communication events.

These improvements mean that the new algorithm can easily process datasets which may have been infeasible in the past. I demonstrate this on a trillion cell DEM. After ruling out ``gargantuan", I follow \citet{Barnes2016} in referring to this new size class as being \textit{rather} large. %Thanks, Miranda!

\section{The Algorithm}
\label{sec:alg_overview}

The algorithm assumes that \textit{non-divergent} flow directions have been previously determined by a separate algorithm of the user's choice. Depressions and flats may or may not be present. The algorithm then efficiently calculates Equation~\ref{equ:flow_acc} based on these flow directions. Since I am considering DEMs which are generally too large to fit into RAM all at once, tiles will be used to calculate \textit{intermediate solutions} which, together, can be used to construct a \textit{global solution}. Although the algorithm is described and implemented in terms of an 8-connected raster, other topologies, such as hexagonal DEMs, could be used.

The algorithm has a single-producer, multiple-consumer design---one process produces tasks, delegates them, and aggregates results, while all the other processes handle the tasks produced---which proceeds in three stages. (1)~The producer allocates tiles to the consumers, which calculate an intermediate based on the tile and pass a small amount of information about the intermediate back to the producer. (2)~Based on this data, the producer calculates the information needed for each consumer to independently produce its share of a global solution. This information takes the form of a flow accumulation offset. (3)~It provides this offset to the consumers, which modify their intermediates based on it. The modified intermediates collectively form the global solution. This design is effectively two sequential MapReduce operations and is general enough to be implemented with either threads or processes using any of a number of technologies including OpenMP, MPI, Apache Spark~\citep{Zaharia2010}, or MapReduce~\citep{Dean2008}. Here, I use MPI.

The third stage of the algorithm modifies intermediates generated by the first stage. But this modification cannot take place until after the second stage has completed. There are three strategies for caching these intermediates which affect both the speed and the memory requirements of the algorithm as a whole. These strategies are as follows. (a)~The \textsc{evict} strategy: a consumer evicts its intermediates from RAM and works on other tiles. This option uses the least RAM and disk space, but requires recalculation of the intermediates later. (b)~The \textsc{cache} strategy: a consumer writes its intermediates to disk in a compressed form (despite the processing requirements, this is faster than storing the data uncompressed~\citep{Barnes2016}) and works on other tiles. \textsc{cache} use the same RAM as \textsc{evict}, but more disk space. Which strategy is fastest will depend on hardware configurations and should be determined by testing. (c)~The \textsc{retain} strategy: a consumer keeps its intermediates in RAM at all times.

If the DEM cannot fit entirely into the RAM of the available node(s), the \textsc{evict} and \textsc{cache} strategies still allow the DEM to be processed. In the limit, only the producer's information and a single tile need be in RAM at a time. \textit{This allows large DEMs to be efficiently processed by a single-core machine}, decreasing resource costs and democratizing analysis. Additional RAM and cores, as may be available on high-end desktops or supercomputers, will result in faster time-to-completion. Only if sufficient RAM is available, such that the entire dataset can be stored in RAM at once, can the \textsc{retain} strategy can be used. This strategy will result in the fastest time-to-completion.

To proceed, the DEM is first subdivided into rectangular tiles of equal dimension. Equal dimensions are not necessary, but requiring it makes the implementation easier. Regardless, this is a natural input format since all of the DEMs considered here are distributed in the form of many square tiles of equal dimension.

\subsection{Solving a Single Tile}
\label{sec:single_block_solve}

Flow accumulations are calculated for each tile separately using any standard serial flow accumulation algorithm. The one I describe here is based on an algorithm by \citet{Wallis2009parallel} and is particularly simple, making it easy to present and verify. Other algorithms~\citep{Braun2013} could be used and may work faster due to better caching properties, though I do not explore this possibility here. If, in the future, even faster algorithms emerge (a route which might be pursued by leveraging GPUs), these could likewise be used.

\begin{figure*}

\includegraphics[width=\textwidth]{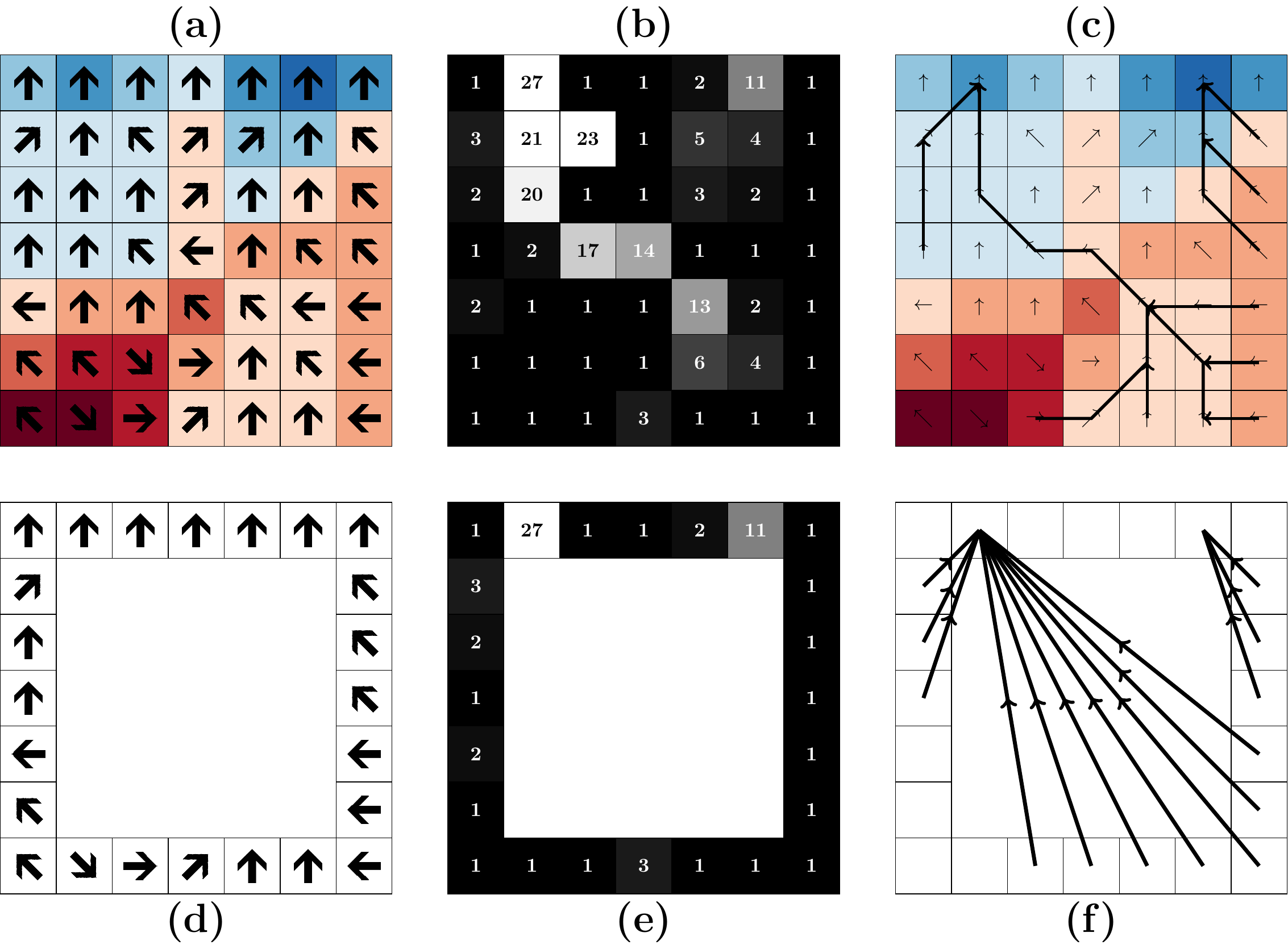}

\hfil%
\resizebox {!} {0.2in} {
\includegraphics[width=\textwidth]{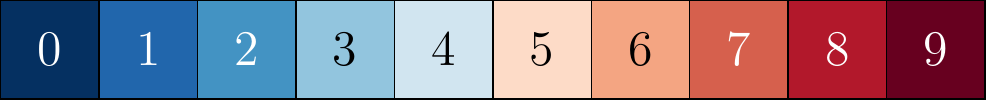}
}%
\hfil%
\caption{Solving a Single Tile. DEM cells are shown as small squares with black borders. Colours in (a) and (c) correspond to elevations, as shown in the legend. Note that elevations are not required by the algorithm: they are included here for explanatory value only. Colours in (b) and (e) correspond to the flow accumulation, with darker shades of grey representing lower values. The flow directions \textbf{(a)} are used to send the perimeter flow directions \textbf{(d)} to the producer. Likewise, once the flow accumulations \textbf{(b)} of the tile have been calculated, the flow accumulations of the perimeter cells \textbf{(e)} are sent to the producer. \textbf{(c)} shows how the perimeter cells are linked together by flow paths. This information is simplified and sent to the producer, as shown in \textbf{(f)}.
\label{fig:single_tile}}
\end{figure*}

To calculate flow accumulation, the new algorithm uses three rasters. (1)~A flow directions raster $F$, which indicates which of a cell's neighbours receives its flow. A cell may also take the special values \textsc{NoFlow} and \textsc{NoData}. \textsc{NoData} denotes a cell which is not part of a DEM, but still contained within its bounding box. \textsc{NoFlow} denotes a cell which is part of a DEM, but without a defined flow direction. Without loss of generality, let us assume here that $F(c)$, rather than being a raw flow direction, is the address of the cell that flow direction points to. (2)~A dependencies raster $D$, which indicates how many of a cell's neighbours flow into it. (3)~A flow accumulation raster $A$, which tracks the total flow passing through each cell. Figure~\ref{fig:single_tile} depicts these arrays graphically. The new algorithm also keeps (4)~a vector $L$ denoting links between perimeter cells. This vector will be explained shortly.

The single tile algorithm proceeds in several stages, which are described below, depicted in Figure~\ref{fig:single_tile}, and shown in Algorithm~\ref{alg:subdiv_fa}.

First, the algorithm begins by initializing $D$ and $A$ to zero.

Second, the algorithm scans through $F$. If, for a cell $c$, $F(c)=\textsc{NoData}$, it is skipped and the corresponding cells $D(c)$ and $A(c)$ are also marked \textsc{NoData}. If $F(c)=\textsc{NoFlow}$, it is skipped. Otherwise, $D(F(c))$ is incremented, provided it is within the tile and not \textsc{NoData}.

Third, any cell with $D(c)=0$ is a cell which receives no flow from any other cell; therefore, its flow accumulation is known: it is simply 1, or any other user-specified weighting value, as in Equation~\ref{equ:flow_acc}. All such cells are collected into a queue $Q$. 

Fourth, a cell $c$ is popped from the queue $Q$. The flow accumulation $A(c)$ of this cell is added to its downstream neighbour $F(c)$, if such a neighbour exists, and the dependency count $D(F(c))$ of this neighbour is decremented. If this results in the neighbour having no dependencies, $D(F(c))=0$, the neighbour, $F(c)$, is added to $Q$.

Fifth, the fourth step is repeated until all cells have been processed. Now every cell's flow accumulation is known, and no cell has any dependencies.

Sixth, the vector $L$ is populated by considering every edge cell in turn. Each edge cell either accepts flow from another tile, passes flow to another tile, both, or neither. To classify a given edge cell $c$, $c$'s flow path ($F(c)$, $F(F(c))$, \ldots) is followed until it either exits the DEM through an edge at some exit cell $e$ or terminates at a \textsc{NoData} or \textsc{NoFlow} cell. Once the termination point of the flow path is found, $L(c)$, is set to either $e$ or the special value \textsc{FlowTerminates}, accordingly. Further, if $F(c)$ itself exits the DEM (has a flow path of length one), then $L(c)$ is set to the special value \textsc{FlowExternal}. Pseudocode for this is shown in Algorithm~\ref{alg:follow_path}.

Cells marked \textsc{FlowTerminates} receive flow from another tile, but do not pass it on to any other tile. They represent the leaves of a \textit{flow graph}. In contrast, cells marked with an exit cell $e$ pass flow through the tile to another point on the perimeter. Cells marked \textsc{FlowExternal} take flow originating within the tile, as well as flow from other tiles, and pass it along to neighbouring tiles.

The fundamental problem each tile encounters is that it does not know how much flow it will receive from each neighbouring tile. Thus, every cell along every flow path is offset from its true value by an unknown amount that may be a function of several variables. Fortunately, this information can be calculated by considering in aggregate the perimeters of all the tiles. Therefore, when a given tile is done being processed as described above, a small amount of information about the tile is sent to the Producer for use in constructing the global solution. This same information is shown in Figure~\ref{fig:single_tile}, in pseudocode in Algorithm~\ref{alg:subdiv_fa}, and with extensively commented supplementary source code.

\begin{breakablealgorithm}
\caption{\small {\sc Single Tile Flow Accumulation}: \textbf{Upon entry,} \textbf{(1)}~$F$ contains the flow direction of every cell or the value \textsc{NoData} for cells not part of the DEM, including those outside the boundaries of the data. In the following, assume that it denotes the address of the cell pointed to by the flow direction. \textbf{(2)}~$F$ may be a tile of a larger DEM. \textbf{At exit,} \textbf{(1)}~$A$ contains the flow accumulation of every cell. $L$ denotes where each perimeter cell links to.}
\label{alg:subdiv_fa}
\footnotesize
\begin{algorithmic}[1]
  \State Let $A$ have the same dimensions as $F$
  \State Let $D$ have the same dimensions as $F$
  \State Set $D(c)=0$ for all $c$
  \State Set $A(c)=0$ for all $c$
  \ForAll{$c$ in $F$} \label{algline:subdiv_fa_loop1}
    \If{$F(c)=\textsc{NoData}$}
      \State $A(c)\gets\textsc{NoData}$
      \State \Continue{\ref{algline:subdiv_fa_loop1}}
    \EndIf
    \If{$F(c)=\textsc{NoFlow}$}
      \State \Continue{\ref{algline:subdiv_fa_loop1}}
    \EndIf
    \If{$F(F(c))=\textsc{NoData}$}
      \State \Continue{\ref{algline:subdiv_fa_loop1}}
    \EndIf
    \State $D(F(c))\gets D(F(c))+1$
  \EndFor
  \State
  \State Let $Q$ be a queue
  \ForAll{$c$ in $D$}
    \If{$D(c)=0$ \textbf{and} $F(c)\ne\textsc{NoData}$}
      \State Add $c$ to $Q$
    \EndIf
  \EndFor
  \State
  \While{$Q$ is not empty} \label{algline:subdiv_fa_loop2}
    \State $c\gets$the top cell of $Q$
    \State $A(c)\gets A(c)+1$
    \If{$F(c)=\textsc{NoData}$}
      \State \Continue{\ref{algline:subdiv_fa_loop2}}
    \EndIf
    \If{$F(c)=\textsc{NoFlow}$ \textbf{or} $F(F(c))=\textsc{NoData}$}
      \State \Continue{\ref{algline:subdiv_fa_loop2}}
    \EndIf
    \State $A(F(c))\gets A(F(c))+A(c)$
    \State $D(F(c))\gets D(F(c))-1$
    \If{$D(F(c))=0$}
      \State Push $F(c)$ onto $Q$
    \EndIf
  \EndWhile
  \ForAll{$c$ on the perimeter of $F$}
    \State $L(c)\gets$\textsc{FollowPath}($c$) \Comment See Alg.~\ref{alg:follow_path}
  \EndFor
\end{algorithmic}
\end{breakablealgorithm}

\begin{algorithm}
\footnotesize
\caption{\small {\sc FollowPath}: For a given perimeter cell $c$ determine which cell it links to.
\textbf{Upon entry,}
\textbf{(1)}~$F$ contains the flow direction of every cell or the value \textsc{NoData} for cells not part of the DEM, including those outside the boundaries of the data. In the following, assume that it denotes the address of the cell pointed to by the flow direction. \textbf{At exit,} \textbf{(1)}~$L$ is updated at position $c$ to denote which perimeter cell $c$ connects to, or to one of the special values \textsc{FlowTerminates} or \textsc{FlowExternal}.}
\label{alg:follow_path}
\begin{algorithmic}[1]
  \State Let $c$ be the perimeter cell in question
  \State $c_0 \gets c$
  \While{\textsc{True}}
    \If{$F(c)=\textsc{NoData}$ \textbf{or} $F(c)=\textsc{NoFlow}$}
      \State $L(c_0)\gets\textsc{FlowTerminates}$
      \State \Return
    \EndIf
    \State $c_n\gets F(c)$
    \If{$c_n$ is outside the tile}
      \If{$c=c_0$}
        \State $L(c_0)=$\textsc{FlowExternal}
      \Else
        \State $L(c_0)=c$
      \EndIf
      \State \Return
    \EndIf
    \State $c\gets c_n$
  \EndWhile
\end{algorithmic}
\end{algorithm}

\subsection{Constructing a Global Solution}
\label{sec:global_solution}

As each tile finishes being processed, as described above, its consumer sends some information about the tile to the producer, as described in the next paragraph. Once this information is sent, the consumer can apply one of the caching strategies described above: \textsc{evict}, \textsc{cache}, or \textsc{retain}. If \textsc{cache} or \textsc{retain} are used, the flow accumulation of each cell must be saved. For \textsc{retain} the flow directions must also be saved.

The consumer sends the following information about each perimeter cell of the tile to the producer: (a)~its flow direction $F$, (b)~its flow accumulation $A$, and (c)~its link information $L$, as shown in Figure~\ref{fig:single_tile}. The amount of information sent is therefore proportional to the length of the tile's perimeter. All of this information is sent only once per tile. Communication costs and data sizes are discussed theoretically in \textsection\ref{sec:complexity} and empirically in \textsection\ref{sec:results} and Table~\ref{tbl:results}.

\begin{figure*}

\includegraphics[width=\textwidth]{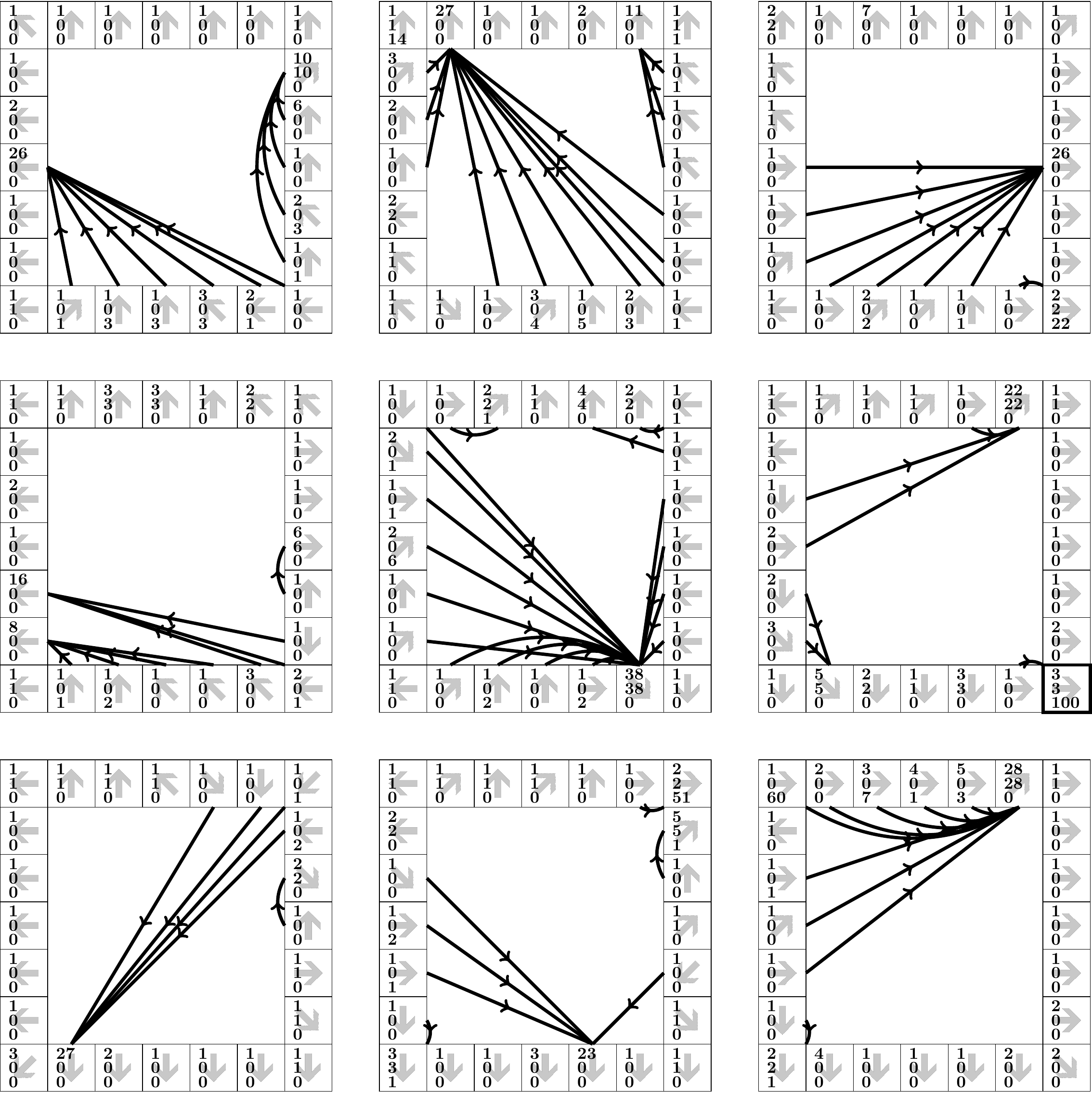}

\caption{The Global Flow Graph. The elevations from which this information is derived are shown in Figure~\ref{fig:stages}. Cells are shown as small squares with black borders and tiles as larger 7\,x\,7 squares separated by white space. Cells' flow directions are indicated by the light gray arrows in each cell. The numbers along the left-hand side of each cell represent, from top to bottom: (a)~its flow accumulation as calculated by the single-tile algorithm (\textsection\ref{sec:single_block_solve}, Algorithm~\ref{alg:subdiv_fa}); (b)~its flow accumulation after cells which are not \textsc{FlowExternal} are set to $A=0$; (c)~its flow accumulation offset $A'$ after the global solution has been calculated. Dark lines with arrows represent links between perimeter cells. The value of 100 in the bolded cell on the middle tile of the right-hand side is achieved by taking inputs from the bottom-right tile ($60+7+1+3+28=99$) and adding 1 for the cell's own addition to its flow accumulation.
\label{fig:producer}}
\end{figure*}

The producer uses non-blocking communication to delegate unprocessed tiles to consumers in round-robin fashion. The producer then uses a blocking receive to collect data from the consumers as they finish processing. Next, each pair of tiles' adjoining edges (or corners) is considered and used to connect the individual tiles together into a single global flow graph, as shown in Figure~\ref{fig:producer}.

The global flow graph is solved using the same strategy described in \textsection\ref{sec:single_block_solve} and Algorithm~\ref{alg:subdiv_fa}, with modifications as follows.
\begin{enumerate}
\item Cells which are not marked as \textsc{FlowExternal} are set to $A=0$ initially. This prevents double-counting: such cells' flow has already been accumulated downstream.
\item Additions to a cell's flow accumulation are tracked in an offset accumulation array $A'$. To pass flow downstream, $A$ and $A'$ are added together. This is because a cell labeled \textsc{FlowExternal} may receive flow from other tiles which must be tracked and which should not be confused with flow originating from within the tile itself.
\end{enumerate}
Finally, the offset matrix $A'$ is distributed to the tiles. Figure~\ref{fig:producer} expands on the foregoing.

\begin{algorithm}
\footnotesize
\caption{\small{ {\sc Main Algorithm}: 
\textbf{Upon entry,}
\textbf{(1)}~\textit{F} contains the flow directions of every cell or the value
\textsc{NoData} for cells not part of the DEM.
\textbf{At exit,}
\textbf{(1)}~\textit{A} contains the flow accumulation of each cell. Communication is assumed to be non-blocking, except where otherwise noted. Consumers perform their calculations asynchronously with respect to the Producer. Note that consumers must be assigned the same tiles in the first and second part of the algorithm for \textsc{retain} to work.}}
\label{alg:main}
\begin{algorithmic}[1]
  \State Let \textit{Consumers} be a thread/process pool
  \State Let a tile have a filename, dimensions, and edge information
  \State Let \textit{Tiles} be a collection of tiles
  \State Let \textit{F} be a flow direction raster
  \State
  \State Divide \textit{F} into tiles
  \ForAll{tiles $b$}
    \State Delegate $b$ to the next consumer $t$
    \State Have $t$ perform Algorithm~\ref{alg:subdiv_fa} on $b$
    \State If there are no more consumers start again at the first
  \EndFor
  \While{any tile is still unreceived}
    \State Block until any consumer returns
    \State Store the information returned
  \EndWhile
  \State
  \State Aggregate the perimeter information into a global flow graph
  \State Generate a flow accumulation offset $A'$ from this flow graph using the modified version of Algorithm~\ref{alg:subdiv_fa}
  \State
  \ForAll{tiles $b$}
    \State Send $b$ and its portion of $A'$ to the next consumer $t$
    \If{$t$ cached the results of Algorithm~\ref{alg:subdiv_fa}}
      \State Let $t$ load the cached results
    \Else
      \State Let $t$ rerun Algorithm~\ref{alg:subdiv_fa}
    \EndIf
    \State Let $t$ add $A'$ to every cell in its downstream flow path
    \State If there are no more consumers start again at the first
  \EndFor
\end{algorithmic}
\end{algorithm}

\subsection{Broadcasting \& Finalizing the Global Solution}
\label{sec:broadcasting}

The foregoing has established the accumulation offsets that need to be added along the flow paths of every tile. Applying this offset is straight-forward: the values from $A'$ are added to every downstream cell of the flow paths they are part of. In order to perform this adjustment, each tile needs the flow accumulations calculated in \textsection\ref{sec:single_block_solve} by Algorithm~\ref{alg:subdiv_fa}. How these are now obtained depends on the chosen caching strategy. If (a)~\textsc{evict} was used, then the intermediate must be recalculated as described by \textsection\ref{sec:single_block_solve}, and then the aforementioned adjustment must be made. Alternatively, if (b)~\textsc{cache} or (c)~\textsc{retain} were used, then a single $O(N)$ sweep of the tile is sufficient to finalize the solution. The pros and cons of these strategies are discussed in \textsection\ref{sec:complexity}.

\begin{figure*}
\hfil%
\resizebox {0.3\textwidth} {!} {
\includegraphics[width=\textwidth]{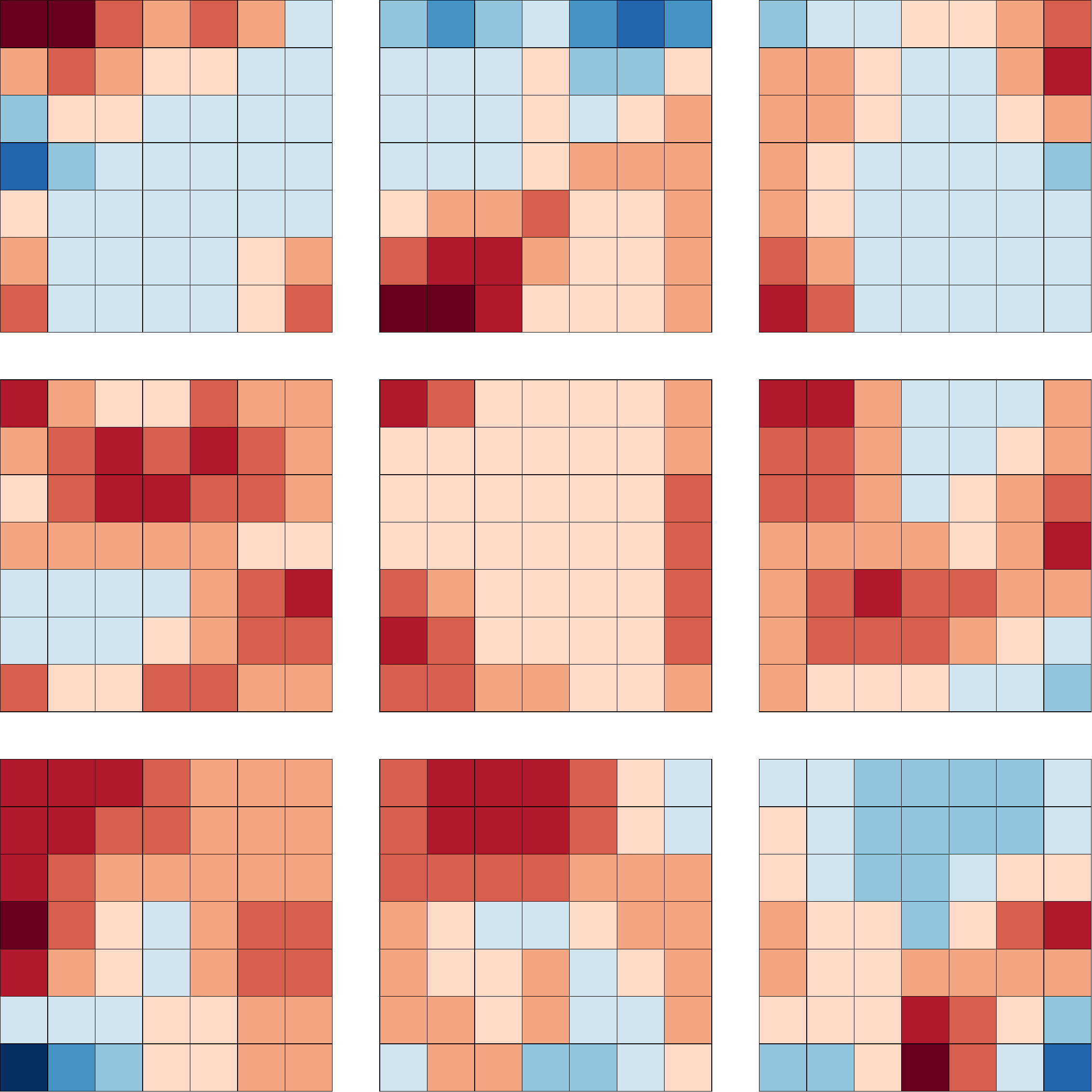}
}%
\hfil%
\resizebox {0.3\textwidth} {!} {
\includegraphics[width=\textwidth]{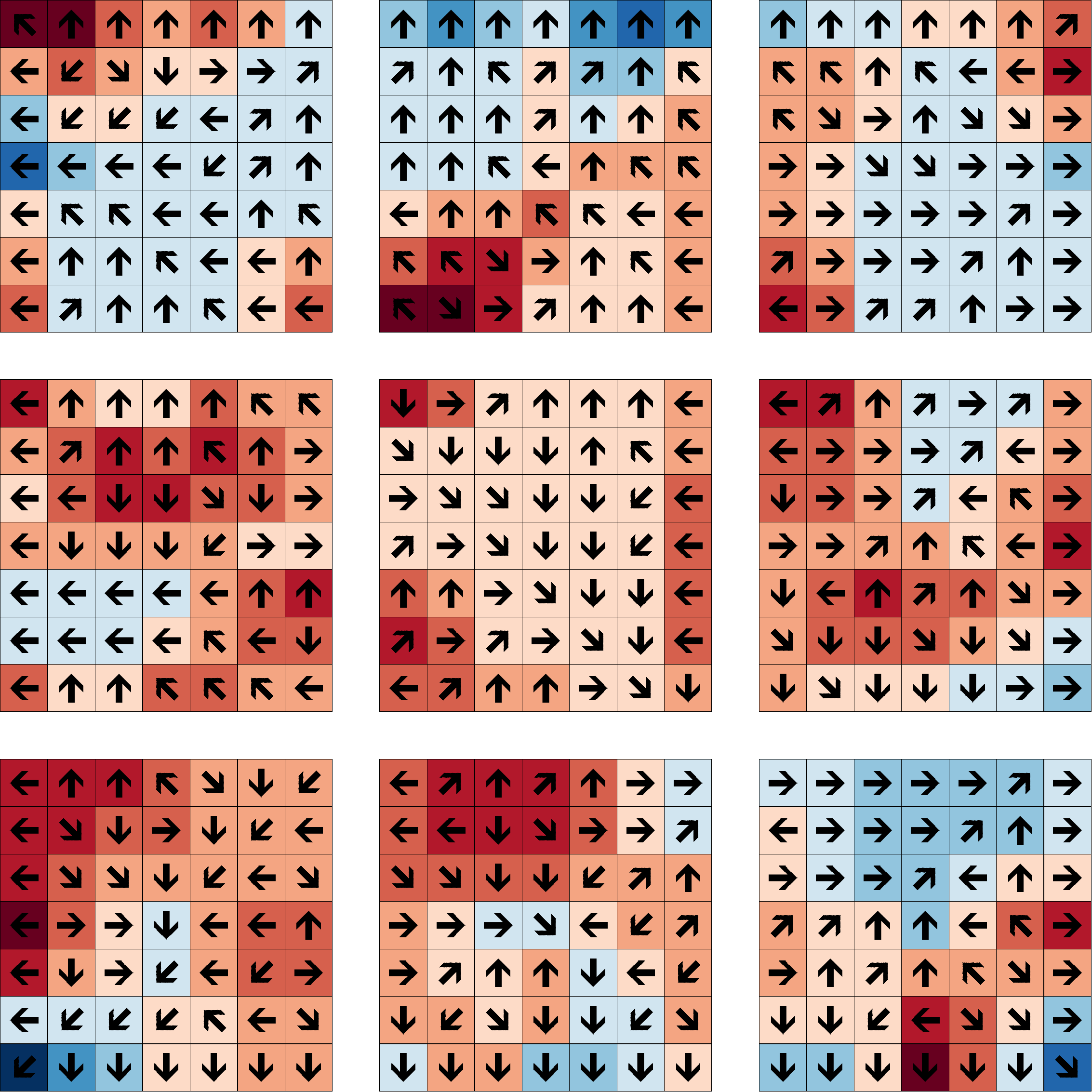}
}%
\hfil%
\resizebox {0.3\textwidth} {!} {
\includegraphics[width=\textwidth]{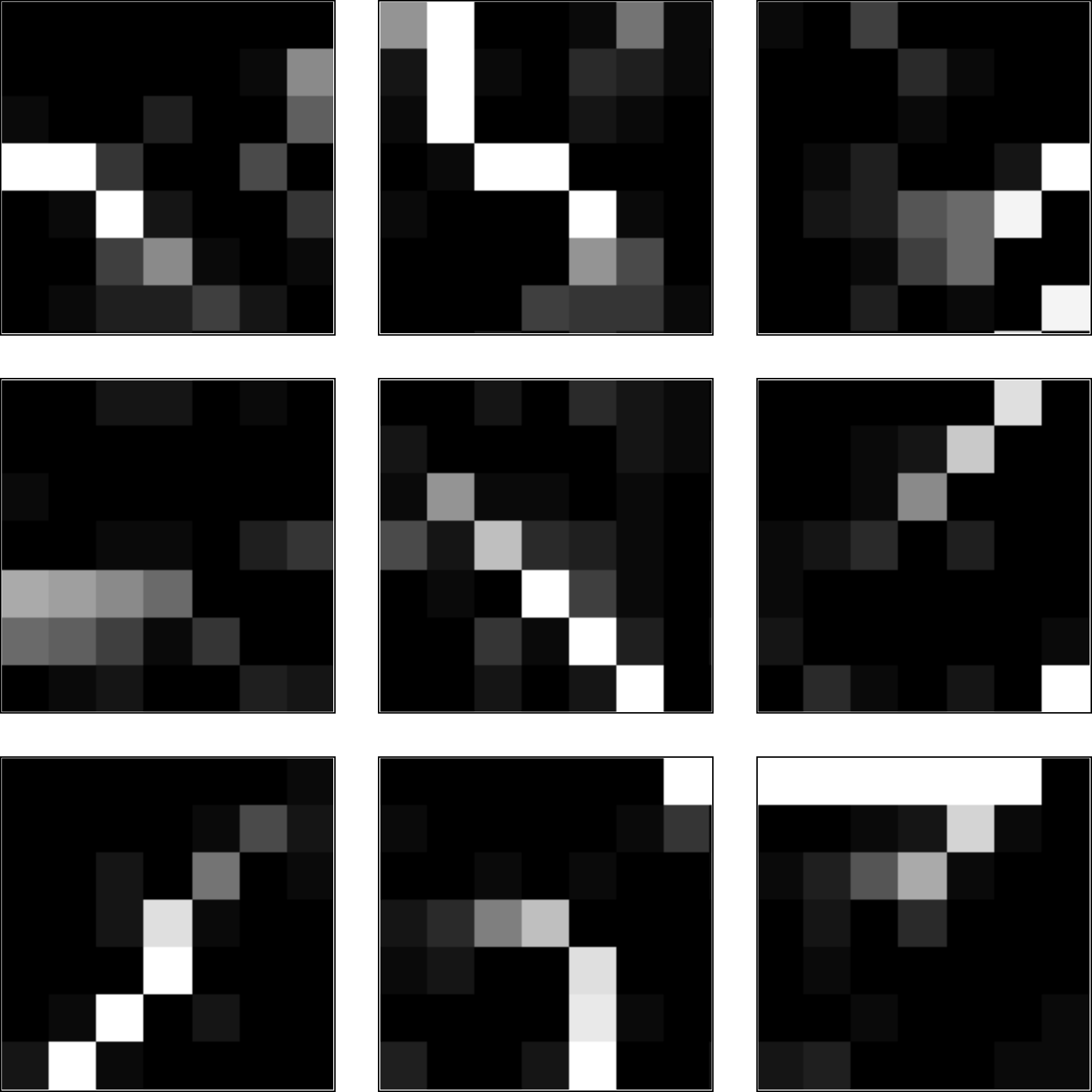}
}%
\hfil%

\vspace{-0.5em}
\hfil a \hfil b \hfil c \hfil

\vspace{0.5em}

\hfil%
\resizebox {!} {0.2in} {
\includegraphics[width=\textwidth]{scale.pdf}
}%
\hfil%
\caption{Overview of the Process. Cells are shown as small squares with black borders and tiles as larger 7\,x\,7 squares separated by white space. Colours in (a) and (b) correspond to elevations, as shown in the legend. Colours in (c) correspond to flow accumulation with darker colours representing less accumulation. Given an elevation raster (a), a separate algorithm treated here as a black box produces flow directions (b). The flow directions are then used to calculate the flow accumulation (c).
\label{fig:stages}}
\end{figure*}

Ultimately, each tile is saved separately to disk for further processing, which may include mosaicing the tiles back into a single, large DEM. The foregoing information is encapsulated in Algorithm~\ref{alg:main} and via extensive comments in the supplementary source code. An overview of the whole process is shown in Figure~\ref{fig:stages}.

\section{Theoretical Analysis}
\label{sec:complexity}

\subsection{Data Types}
In the new algorithm, the data type of the flow directions is fixed at
1~byte/cell. This is sufficient to indicate which of 8 neighbours a cell should point to, as well as to indicate \textsc{NoFlow} and \textsc{NoData}. Space could be saved by packing these ten distinct values into a nibble, but I do not pursue this optimization here.

For the largest raster considered, the worst-case flow accumulation value is ${\sim}10^{12}$. Since GDAL does not use 64-bit integers, the double-precision floating-point data type is used to measure accumulation. The IEEE754 standard specifies that this type will have a 53-bit significand. Therefore, exact integer precision can be maintained for datasets up to ${\sim}10^{16}$ cells. This should be sufficient for most current and future applications, though the compromise on precision could be easily remedied through modifications to GDAL.

The Links array must be able to address any cell on the perimeter of a tile. Therefore, its data type must be able to hold values at least as long as this perimeter. An unsigned 16-bit integer should be suitable for this as it allows values up to 65,535, which permits tiles of ${\sim}16384^2$.

\subsection{Time Complexity}
The time complexity of the algorithm is a function of the time taken to process each individual tile and the time taken to build the global solution. Individual tiles are processed using a serial flow accumulation algorithm. If $n$ is the number of cells per tile, all such algorithms take $O(n)$ time per tile. Variations in the run-time of these algorithms will be dominated by cache behavior.

The global solution requires that flow accumulation be performed on the global flow graph aggregated from the tiles' perimeters. The number of nodes in this graph is exactly equal to the number of cells in all the tiles' perimeters. A single tile has a perimeter of ${\sim}4\sqrt{n}$ cells. If we call the number of tiles $T$, then the global solution takes $O(T\sqrt{n})$.

Once this graph has been processed, if the individual tiles were cached, then at worst $O(n)$ accumulation offsets must be applied; otherwise, if \textsc{evict} was used, the flow accumulation must be performed on each tile followed by the application of offsets. Either way, finalizing takes $O(n)$ time.

Therefore, in the worst-case, the total time is $O(Tn)$. This is divided evenly between the available processors giving a total time of $O(Tn/p)$. Running a flow accumulation on the entire dataset at once would take $O(Tn)$ time. Therefore, the new algorithm is faster than a serial algorithm in proportion to the number of available cores. Thus, we do not expect a significant reduction in raw processing time. However, the new algorithm's disk access and communication patterns, discussed below, represent a significant theoretical and observed speed-up over previous approaches.

\subsection{Disk Access}

The new algorithm guarantees that each tile, and therefore, each cell, need only be loaded into memory a fixed number of times. Recall from \textsection\ref{sec:alg_overview} that there are three memory retention strategies. (a)~\textsc{retain}. The entire dataset is retained in the memory of the nodes at all times: this requires one read and one write per cell. (b)~\textsc{cache}. The dataset cannot fit entirely into the memory of the nodes, so intermediate results (flow accumulations) are cached to disk: this requires less than three reads and two writes per cell when compression is used. (c)~\textsc{evict}. No intermediates are cached: this requires two reads and one write per cell.

\textsc{retain} is the fastest strategy, but unlikely to be feasible for large datasets. \textsc{cache} reduces computation versus \textsc{evict}, but is more expensive in terms of disk access. With compression, \textsc{cache} may use nearly any amount of computation depending on the algorithm employed: a good algorithm should yield acceptable compression with minimal processing. Previous algorithms based on virtual tiles must be at least as expensive as \textsc{retain}. Each time such an algorithm swaps a virtual tile out of memory, it incurs the cost of one write (and, later), one read. Therefore, if approximately half the virtual tiles are swapped once, the costs will surpass \textsc{evict}. Put another way: if the dataset is twice as large as the available RAM, it is reasonable to expect a virtual tile algorithm to be more expensive than that presented here. Given the size of the test sets I employ, this is almost certainly the case.

\subsection{Communication}

Disregarding data structure overhead, the new algorithm needs to pass information regarding $F$, $A$, and $L$ for each of the tile's $4\sqrt{n}$ edge cells to the producer at a cost of $(4\sqrt{n})(1+8+2)$ bytes. In turn, for each tile the producer passes back \textsc{AccumOffset} for each edge cell at a cost of $(4\sqrt{n})(8)$. Therefore, the total communication cost is approximately $(4\sqrt{n})(19)$ per tile.

Previous parallel implementations have exchanged edge accumulation information
between adjacent tiles after each iteration of their algorithms. For a tiled
dataset, the cost between two cores is $(2\sqrt{n})(8)$ bytes per iteration.
Therefore, the cost of communication between two cores in a previous algorithm
surpasses the cost of communication between a producer and a single consumer in
the new algorithm after five iterations. Again, for a large dataset this is easily exceeded.

%cat main.cpp ../../include/richdem/common/Layoutfile.hpp ../../include/richdem/common/communication.hpp ../../include/richdem/common/memory.hpp ../../include/richdem/common/timer.hpp ../../include/richdem/common/Array2D.hpp ../../include/richdem/common/grid_cell.hpp | d.removeblanks | grep -E '^\s*//|^\s*\*' | wc

%cat main.cpp ../../include/richdem/common/Layoutfile.hpp ../../include/richdem/common/communication.hpp ../../include/richdem/common/memory.hpp ../../include/richdem/common/timer.hpp ../../include/richdem/common/Array2D.hpp ../../include/richdem/common/grid_cell.hpp | d.removeblanks | wc

\section{Empirical Tests}

\label{sec:tests}

I have implemented the algorithm described above in \texttt{C++11} using MPI for communication, the Geospatial Data Abstraction Library (GDAL)~\citep{GDAL} to read and write data, Cereal to serialize data during communication~\citep{Grant2013}, and Boost IOstreams to handle compression for the \textsc{cache} strategy. Tests were performed using Intel MPI v5.1; the code is also known to work with OpenMPI v1.10.2. There are 2,131 lines of code of which 58\% are or contain comments.\todo{Check again before publishing} Since the algorithm does not rely on details of the communication, implementing the algorithm with Spark or MapReduce or would be straight-forward. The code can be acquired from \url{https://github.com/r-barnes/Barnes2016-ParallelFlowAccum}.

\newcolumntype{R}[2]{%
    >{\adjustbox{angle=#1,lap=\width-(#2)}\bgroup}%
    l%
    <{\egroup}%
}
\newcommand*\rot{\multicolumn{1}{R{45}{1em}}}% no optional argument here, please!

\begin{table*}
\footnotesize
\centering
\begin{tabular}{l l l l l lll S[table-format=3.2] S[table-format=3.2] l S[table-format=3.2]}
{DEM}         & {Resolution}     & {Tiles}     & {Cells/Tile}     & {Tile Size}    & {Total Size}     & {Cells}       \\ \hline
%SRTM Resampled & 10\,m            & 14297       & 10803$^2$        & 233\,MB           & 3.34\,TB         & $1.7\cdot10^{12}$ \\
SRTM Global    & 30\,m            & 14297       & 3601$^2$         & 13\,MB             & 185\,GB          & $1.9\cdot10^{11}$  \\
NED            & 10\,m            & 1023        & 10812$^2$        & 117\,MB            & 120\,GB          & $1.2\cdot10^{11}$  \\
PAMAP North    & 1\,m             & 6666        & 3125$^2$         & 9.7\,MB            & 65\,GB          & $6.5\cdot10^{10}$  \\
PAMAP South    & 1\,m             & 6723        & 3125$^2$         & 9.7\,MB            & 66\,GB          & $6.6\cdot10^{10}$  \\
SRTM Region 1  & 30\,m            & 164         & 3601$^2$         & 13\,MB             & 2.1\,GB          & $2.1\cdot10^{ 9}$  \\
SRTM Region 2  & 30\,m            & 161         & 3601$^2$         & 13\,MB             & 2.1\,GB          & $2.1\cdot10^{ 9}$  \\ \hline
\end{tabular}
\caption{Datasets employed for testing the new algorithm.
\textbf{Tiles} indicates the number of tiles the DEM was divided into by its
provider. \textbf{Tile Size} indicates how much uncompressed space it would take
to store the number of cells in the tile, given its data type (cell count times
data type size: in this case, one byte). \textbf{Total Size} indicates how much space it would take to
store all of the tiles in the dataset.
\label{tbl:my_datasets}}
\end{table*}

To demonstrate the scalability and speed of the algorithm, I tested it on several large DEMs, including one \textit{rather} large one, as shown in \autoref{tbl:my_datasets}. All of these DEMs came pre-divided into equally-sized tiles by their providers; I used these existing tile structures in most of my tests.

The DEMs tested include
\begin{itemize}
\item PAMAP\footnote{\url{ftp://pamap.pasda.psu.edu/pamap_LiDAR/cycle1/DEM/}}: A LiDAR
DEM covering the entire state of Pennsylvania. The data is available as 13,918
tiles divided into a north section and a south section. These sections are
projected differently and, therefore, the two are considered independently here.
%At a resolution of 0.98\,m (3.2\,ft).%The PAMAP LiDAR elevation data was
%collected from 2006--2008 with 1.4\,m average point spacing and a vertical
%accuracy at least 18.5\,cm (RMSE) in open areas.

\item NED\footnote{\url{ftp://rockyftp.cr.usgs.gov/vdelivery/Datasets/Staged/Elevation/13/IMG/}}:
National Elevation Dataset 10\,m data. Higher resolution 3\,m and 1\,m data are
available, but only in patches, whereas 10\,m data are available for the entire
conterminous United States, Hawaii, and parts of Alaska. The entire 10\,m NED
DEM is considered here as a single unit. Although islands are present in the
DEM, the algorithm implicitly handles these without an issue.
%The mean absolute vertical accuracy of the dataset is
%$<$1\,m.~\citep{Gesch2014} 

\item SRTM:
Shuttle Radar Topography Mission (SRTM) 30\,m DEM. This 30\,m data covers 80\%
of Earth's landmass between 56$^\circ$S and 60$^\circ$N. The data was
originally available as several regions covering North
America\footnote{\url{http://dds.cr.usgs.gov/srtm/version2_1/SRTM1/}}, which
are considered separately here; more recently, global
data\footnote{\url{http://e4ftl01.cr.usgs.gov/SRTM/SRTMGL1.003/2000.02.11/}}
has been released. The global data is considered as a single unit here. Since the surfaces of oceans and the like are topographically uninteresting, tiles which would contain only oceans are not present in the dataset.
\end{itemize}

Further details on acquiring the aforementioned datasets are available with the
source code.

%The data was collected
%in February 2000 by the Space Shuttle Endeavour and mapped Earth's topography
%between 56$^\circ$S and 60$^\circ$N (80\% of the Earth's landmass) over an
%eleven day period using an imaging radar. Although 30\,m data was released for
%several regions of the United States shortly thereafter, only 90\,m data was
%released for other parts of the world. Beginning in 2014, the entire dataset was
%gradually released at 30\,m. The data have an absolute accuracy of $\le$16\,m.
%Here, each U.S.\ region is considered independently and the entire global
%dataset is considered as a single unit.

Tests were run on the Comet machine of the Extreme Science and Engineering Discovery Environment (XSEDE)~\citep{xsede}. Each node of the machine has 2.5\,GHz Intel Xeon E5-2680v3 processors with 24 cores per node, 128\,GB of DDR4 DRAM, and 320\,GB of local SSD storage. Nodes are connected with 56\,Gbps FDR InfiniBand. Data were held in Oasis: a 200\,GB/s distributed disk Lustre filesystem. Code was compiled using GNU g++ 4.9.2.

Four tests were run. For each test, the new algorithm was run using the \textsc{evict} strategy to simulate a minimal-resource environment.

The first test ran the algorithm on two nodes (48 cores) for each of the datasets listed in \autoref{tbl:my_datasets} using the full dataset and all of the available cores. The result is shown in \autoref{tbl:results}.

All of the datasets contain islands of data surrounded by empty tiles, or have irregular boundaries. Therefore, in order to test scaling, the largest square subset of contiguous tiles was identified in each dataset. The resulting subsets were 44\,x\,44 (PAMAP North and South), 39\,x\,39 (SRTM Global), 19\,x\,19 (NED), 11\,x\,11 (SRTM Region 1 and 2).

\begin{figure*}
\centering
\begin{tabular}{cc}
\includegraphics[width=0.45\textwidth]{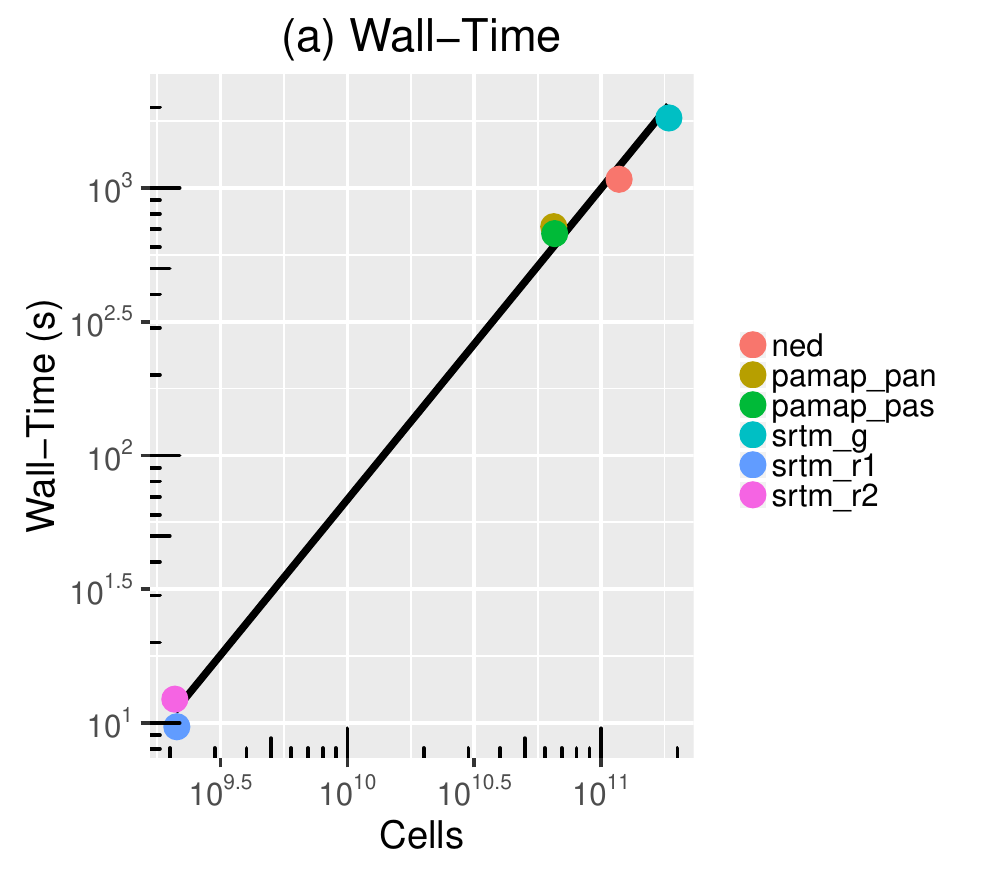} &
\includegraphics[width=0.45\textwidth]{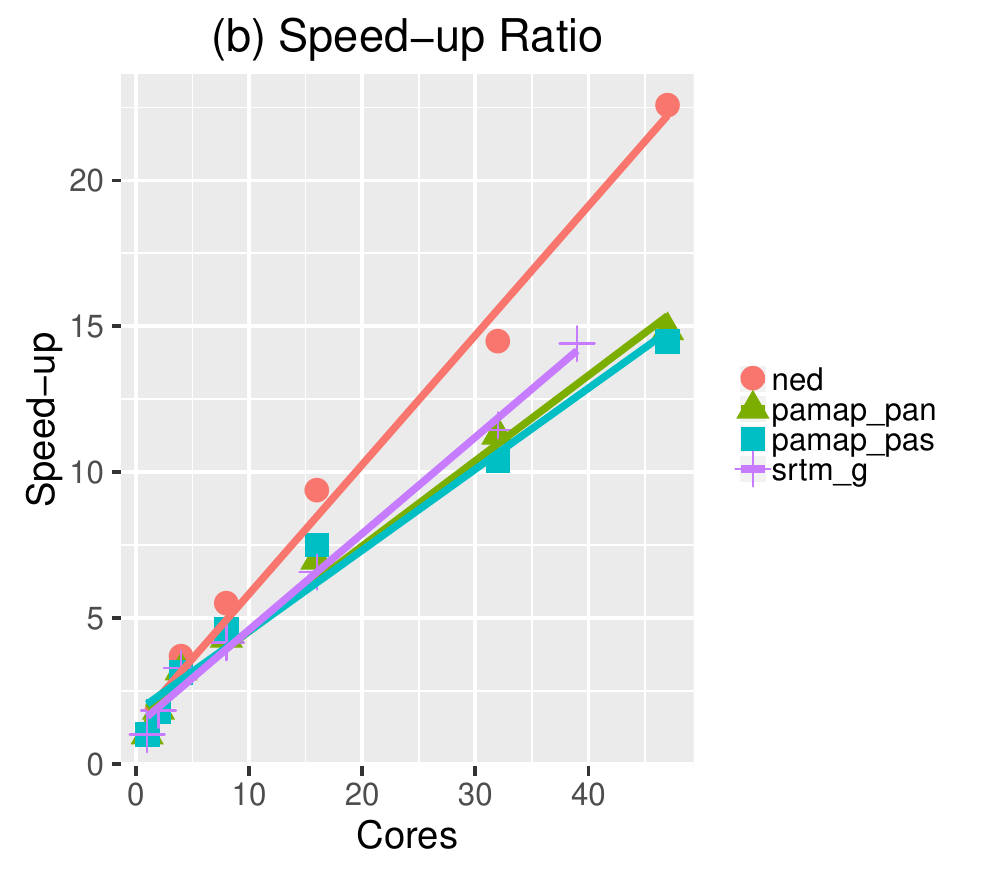} \\
\includegraphics[width=0.45\textwidth]{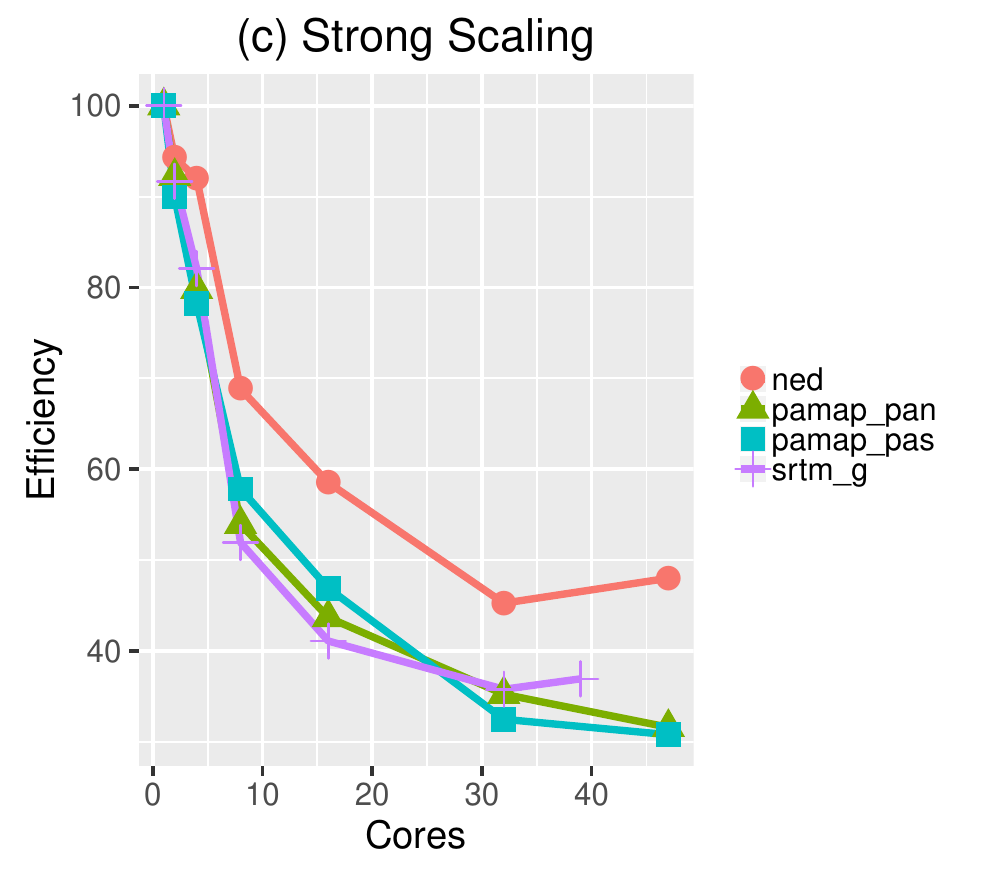} &
\includegraphics[width=0.45\textwidth]{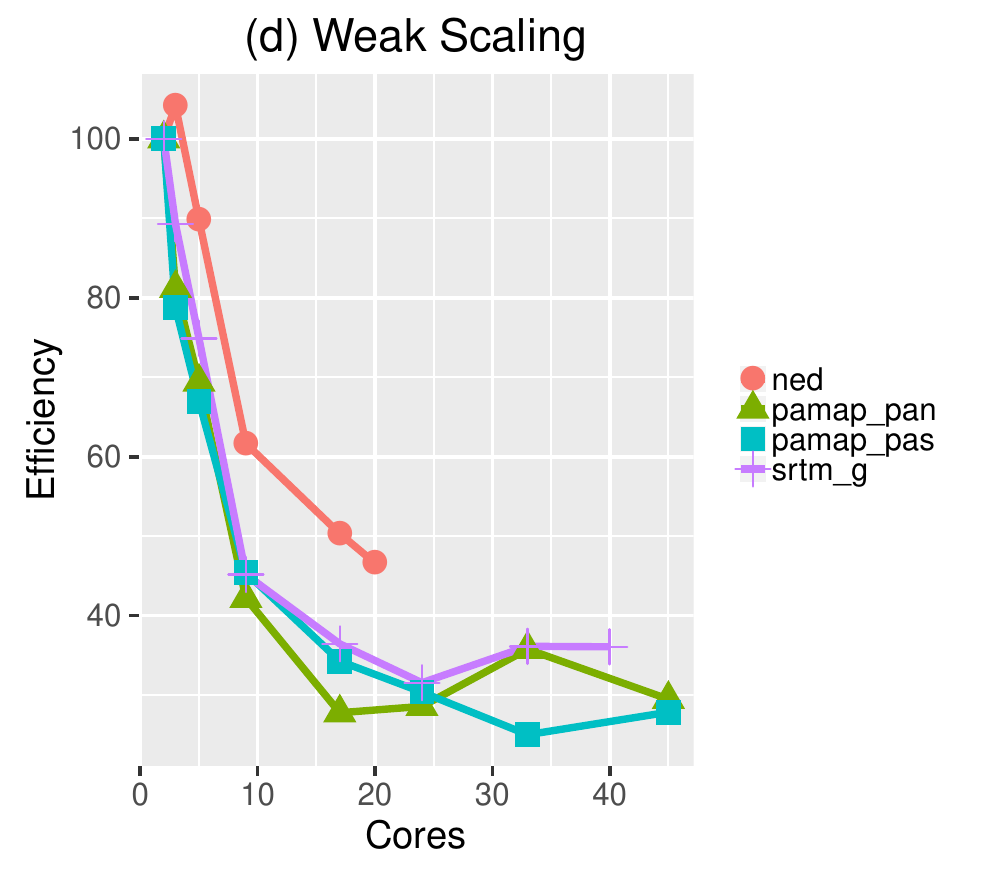}
\end{tabular}
\caption{Results. Let $N$ be the number of cores used, $t_1$ be the time taken by one core to perform one work unit, and $t_N$ be the time taken by $N$ cores to perform the job. The speed-up ratio is given as $\frac{t_1}{t_N}$ where the job size is unchanged. Strong scaling is given by $\frac{t_1}{N t_N}$ where the job size is unchanged. Weak scaling is given by $\frac{t_1}{t_N}$ where the job size is increased proportionally to $N$. In \textbf{(a)} 48 cores are used. \label{fig:result_graphs}}
\end{figure*}

The second and third tests were performed on these contiguous square subsets.
Strong scaling efficiency is a metric of an implementation's ability to solve a
problem faster by using more resources. To test this, increasing numbers of
cores (up to 48) were used on the full square subsets. Weak scaling efficiency
is a metric of an implementation's ability to solve proportionately larger
problems in the same time using proportionately more resources. To test this,
one core was used to process one row of each square subset, two cores for two
rows, and so on. The results are shown in \autoref{fig:result_graphs}.

In a fourth test, a comparison was made against the work of both \citet{Wallis2009parallel} (TauDEM\footnote{\texttt{e19dc083e}, master, \url{https://github.com/dtarb/TauDEM}}) and \citet{Gomes2012} (EMFlow\footnote{\texttt{0ca9e0ef0}, master, \url{https://github.com/guipenaufv/EMFlow}}). These algorithms have source code available, claim to be suitable for large datasets, and claim to be faster than other algorithms, including ArcGIS.

To handle the input limitations of EMFlow, a 40,000\,x\,40,000 single-file DEM was constructed by merging SRTM Region 2 data. All the packages were compiled using GNU g++ 4.9.2 with optimizations enabled. \texttt{/usr/bin/time} and \texttt{mpiP}\footnote{\url{http://mpip.sourceforge.net}} were used to measure memory usage as well as communication times and loads. Both attach to programs at runtime, eliminating the need for modification.

Since EMFlow is single-threaded, the new algorithm, TauDEM, and EMFlow were compared using a single active core. Additionally, the new algorithm and TauDEM were compared using 24 cores distributed across a single node.

\begin{table*}
\footnotesize
\centering
\begin{tabular}{l llll lll l l l l}
                & \rot{Time}       & \rot{Sec/$10^9$ cells}  & \rot{\% I/O} & \rot{All Time}       & \rot{Prod.\ Calc}    & \rot{Sent}        & \rot{Received}    & \rot{Tx/Tile}         & \rot{Cons.\ VmHWM}         & \rot{Prod.\ VmPeak} \\                  
DEM             & Min              &                         &  \%          & Hrs                  & Sec                  & MB                & MB                & KB                    & MB                         & MB                  \\ \hline
SRTM Global     & 24.0             & 7.8                     & 64           & 14.5                 & 19                   & 1,651             & 2,263             & 274                   & 258                        & 6,424               \\
NED             & 14.8             & 7.6                     & 55           & 7.1                  & 3.7                  & 349               & 480               & 822                   & 1,975                      & 1,393               \\
PAMAP North     & 9.6              & 8.9                     & 61           & 5.2                  & 7.3                  & 669               & 917               & 238                   & 199                        & 2,915               \\
PAMAP South     & 9.6              & 8.8                     & 65           & 5.7                  & 7.5                  & 675               & 925               & 238                   & 199                        & 2,936               \\
SRTM Region 1   & 0.4              & 12.3                    & 51           & 0.1                  & 26                   & 19                & 26                & 274                   & 239                        & 382                 \\
SRTM Region 2   & 0.5              & 13.4                    & 64           & 0.1                  & 28                   & 19                & 26                & 274                   & 239                        & 382                 \\ \hline
\end{tabular}
\caption{Results.
\textbf{Time} is the time-to-completion (aka wall-time) of the
algorithm. \textbf{Sec/$10^9$ cells} indicates how many wall-time seconds it
took the algorithm to process a billion cells on each dataset. \textbf{All Time}
indicates the sum of the processing and I/O time of every CPU core used by the
algorithm; this is the unit by which supercomputing centers charge. \textbf{\% I/O}
indicates what percentage of the All Time value was spent on reading and writing
data. \textbf{Prod.\ Calc} is the amount of time the producer spent calculating
the global solution. \textbf{Sent} is the amount of data sent by the producer.
\textbf{Received} is the amount of data received by the producer.
\textbf{Tx/Tile} is the sum of the data received and sent divided by the number
of tiles in the dataset. \textbf{Cons.\ VmHWM} is the virtual memory ``high
water mark" used by one of the consumers to store its data, as determined by the
Linux kernel. \textbf{Prod.\ VmPeak} is the peak virtual memory used by the
producer to store its data and the shared libraries it uses, as determined by
the Linux kernel.
\label{tbl:results}}
\end{table*}

% \begin{table}
% \footnotesize
% \centering
% \begin{tabular}{lcccc}
% DEM           & \textsc{evict} & \textsc{cacheC} & \textsc{retain} \\ \hline %TODO: Check
% SRTM Region 1 & 26.9           & 76.2   (0.4\,x) & 23.5   (1.1\,x) \\
% SRTM Region 2 & 27.9           & 76.8   (0.4\,x) & 24.2   (1.2\,x) \\
% \end{tabular}
% \caption{Timing results in seconds, and speed-up factors versus \textsc{evict}, for different caching strategies. \label{tbl:strategy_comparison}}
% \end{table}

%Variation in this value is driven in part by different data formats.

\section{Results \& Discussion}
\label{sec:results}

\subsection{Comparisons}

In \textsection\ref{sec:alt_algs} I argued that the new algorithm should scale better than existing algorithms because it can use multiple cores, and has fixed I/O and communication requirements. The results of my tests support this.

EMFlow running with a maximum of 2\,GB RAM and tiles of 400\,x\,400 cells (the settings discussed by \citet{Gomes2012}) had 658\,s wall-time and used 2.2\,GB RAM. Running with one process, TauDEM took 530\,s and used 20.4\,GB RAM. In contrast, the new algorithm working on tiles of 4,000\,x\,4,000 cells gave 152\,s wall-time and used 0.4\,GB RAM.

On the 40,000\,x\,40,000 test set, TauDEM with 24 processes had 42\,s wall-time, transmitted 556\,MB, used 1,256\,s for communication, and took 21.1\,GB RAM. The new algorithm (running with a tile size of 4,000\,x\,4,000) had 23.9\,s wall-time, transmitted 30\,MB, used 163\,s for communication, and took 6.7\,GB RAM. Communication time is greater than wall-time because it is a summation across many cores. The foregoing confirms the predictions made in \textsection\ref{sec:complexity}.

Based on the foregoing, I conclude that in all respects the new algorithm is both faster and uses fewer resources than existing algorithms, at least for datasets of the size tested.

\subsection{Flexible Operation}

The above demonstrates that the algorithm can leverage many-core systems, but also operate well with much more limited resources. \autoref{tbl:results} provides further confirmation of this. VmPeak shows the maximum RAM used by the producer to hold both its data and the shared libraries used by the program, and VmHWM shows the maximum RAM used by a consumer. Since the producer and consumers trade off operation, they do not contend for computational resources. Therefore, the memory required to process a DEM using only one consumer is approximately the sum of VmHWM and VmPeak: 6.7\,GB for a 185\,GB dataset in the largest case. The compute time required for such an operation is given by the ``All Time" column of \autoref{tbl:results}, since the time required for calculations by the producer is negligible (19\,s in the largest case).

%As \autoref{tbl:strategy_comparison} shows, running the algorithm's various strategies on the SRTM regional data provides further evidence of the algorithm's flexibility. While the \textsc{cache} strategy does not seem to provide a performance advantage, the \textsc{cacheC} strategy saves several seconds of processing time. On larger datasets, this could make a noticeable difference. Clearly, when resources are available, utilizing the \textsc{retain} strategy is worthwhile.

%If fewer resources are available, the algorithm the tiles can be divided
%between a limited number of cores. In the most extreme case, tiles can be
%processed one at a time. This means the architecture of the machine the
%algorithm runs on does not affect its ability to run: large datasets can still
%be processed efficiently. The algorithm also offers several memory retention
%strategies (mentioned above) to help it adapt to different environments. This
%contrasts with previous algorithms which, in many cases, cannot run unless they
%are matched to an appropriate architecture.

\subsection{Scaling}

In \textsection\ref{sec:complexity}, I argued that the algorithm should scale linearly with the number cells for a fixed tile size. \autoref{fig:result_graphs}a shows this to be partly true: a linear fit to the log-log plot has a slope of 1.2 (R$^2=1.00$) across datasets whose sizes differ by three orders of magnitude. The deviance from 1.0 is likely due to memory effects that are not well-captured by the time complexity analysis.

Figures \ref{fig:result_graphs}c and \ref{fig:result_graphs}d show sustained efficiencies of 30\% on up to 48 cores distributed across two nodes for the datasets with smaller tiles. The larger tiles of the NED result have higher efficiencies of 50\%, likely due to lower I/O overhead. As a result, as the number of cores increases, the speed-up ratio shown in Figure~\ref{fig:result_graphs}b is approximately linear with an average slope of 0.34 across all datasets. %This contrasts with the results of \citet{Yildirim2015} whose implementation quickly reached diminishing returns (see their Figure 7).

\subsection{Larger datasets}

Can even larger, perhaps even \textit{unusually} large, datasets be used? Yes. No fundamental limit prevents the algorithm from scaling to even larger datasets than those tested here. As Figure~\ref{fig:result_graphs} shows, the algorithm's time complexity is approximately linear and it scales decently across large numbers of cores. Additionally, the processing time required by the producer is negligible in comparison to the total, and the per-tile communication requirements are low. The 6.7\,GB RAM and 14.5 compute-hours required for the SRTM-G dataset are well within the limits of a high-spec laptop.

A more complex implementation could reduce the producer's memory requirements by performing partial computation of the global solution as tiles return their data. For clarity, I have opted to build a simpler implementation which stores all of the tiles' returned data in memory prior to calculating the global solution.

\subsection{Speed improvements} The algorithm can run faster. Although the flow accumulation algorithm presented here is optimal in terms of time complexity, its run-time is dominated by its cache behaviour. My personal experiments with an algorithm by \citet{Braun2013} suggest that there may be ways to ameliorate this. It may also be possible to leverage GPUs for greater speed. However, Table~\ref{tbl:results} makes it clear that the greatest gains will come from optimizing I/O. Possible methods include using the Lustre filesystem's stripe option, prefetching data to nodes' SSDs, and utilizing GeoTIFF's data compression options. However, the efficacy of these optimizations will be dependent on the architecture of the test system, so I do not pursue them here.

\subsection{Robustness}

The algorithm is robust in the face of crashes and other interruptions. The data each tile sends to the central node could be cached, allowing the
algorithm to proceed without having to repeat work after a crash. Once the
central node has calculated a global solution, this solution can be cached and loaded in order for finalization to continue. For simplicity, I have not yet included this capability in my own implementation.

\subsection{Correctness}

A formal proof of correctness is beyond the scope of this paper. However, I have built an automated tester which performs correctness tests on arbitrary inputs. This tester, along with several tests, is included in the source code.

In any test, a correct result must be established. While ArcGIS or GRASS could be used for this, doing so would introduce a large and potentially expensive dependency that could not be included with the source code. Therefore, I use a simple implementation of a flow accumulation algorithm to establish correct results. This algorithm can be verified by inspection and then used to test the new algorithm.

In testing, a dataset's tiles are merged using \textsc{gdal} and treated as a single unit to generate an authoritative answer. The new algorithm is then run on the uncombined tiles. In all cases, the algorithm is run with each of its memory retention strategies. Running this suite of tests on a number of inputs did not show any deviation from the authoritative answer, which is evidence of correctness.

\section{Coda}

A limitation of the algorithm presented here is that it performs only simple flow accumulation while there are many other properties that may be of interest. \citet{Tarboton2008} suggest the possibility of a generalized ``flow algebra" which could capture the calculations of these many properties in a single system. In future work, I will investigate generalizing the techniques presented here and in \citet{Barnes2016} for use with flow algebras to form a very general approach for extracting hydrological features and properties from DEMs. Another limitation is that the new algorithm requires that the flow metric it considers be non-divergent. In future work I will endeavour to relax this limitation.

%Could be easily extended to rho8 fairfield 1991
%MFD, Freeman 1991 harder
%Dinfty, Tarboton  harder

In summary, prior flow accumulation algorithms for large digital elevation
models required massive centralized RAM, suffered from unpredictable and slow
disk access when a virtual tile approach was used, or required large numbers of
nodes and communications when parallel processing was used. In contrast, the
present work has introduced a new algorithm which ensures fixed numbers of disk
accesses and communication events. This enables the efficient processing of
\textit{rather} large DEMs constituting trillions of cells on both high- and low-resource machines.

Complete, well-commented source code, an associated makefile, and correctness tests are available at \url{https://github.com/r-barnes/Barnes2016-ParallelFlowAccum}. This algorithm is part of the RichDEM (\url{https://github.com/r-barnes/richdem}) terrain analysis suite, a collection of state of the art algorithms for processing large DEMs quickly.

\section{Acknowledgments}
Early-stage development utilized supercomputing time and data storage provided by the University of Minnesota Supercomputing Institute. Empirical tests and results were performed on XSEDE's Comet supercomputer~\citep{xsede}, which is supported by the National Science Foundation (Grant No.\ ACI-1053575).

Funding: This work was supported by the National Science Foundation's Graduate Research Fellowship and by the Department of Energy's Computational Science Graduate Fellowship (Grant No.\ DE-FG02-97ER25308).

\section{Bibliography}

{\footnotesize
  \bibliographystyle{elsarticle-harv} %TODO: Doesn't seem to be needed for upload
  \bibliography{refs}   % expects file "myrefs.bib"
}

\end{document}